\documentclass[manuscript]{aastex}

\def\gta{\gtrsim}
\def\lta{\lesssim}

\def\ms{{M_*}}
\def\be { \begin{equation} }
\def\ee { \end{equation} }
\def\L {{\cal L}}
\def\LLp {{ {m_1\over\mp}\left({j_1\over\jp}\right)^{1/3}
\sqrt{1-e_1^2\over 1-\ep^2} }}
\def\jp {{j_p}}
\def\np {{n_p}}
\def\mp {{m_p}}
\def\ep {{e_p}}
\def\ap {{a_p}}
\def\Ep {{E_p}}
\def\Lp {{L_p}}
\def\Yp {{Y_p}}
\def\dEdEp {{1+{m_1\over \mp}\left({\jp\over j_1}\right)^{2/3} }}

\begin{document}


\title{\bf Eccentricity Evolution of Resonant Migrating Planets} 
\author{N.  Murray\altaffilmark{1},
M. Paskowitz, and M. Holman\altaffilmark{2}}

\altaffiltext{1}{Canadian Institute for Theoretical Astrophysics,
 60 St. George st., University of Toronto, Toronto, ONT M5S 3H8,
Canada; murray, paskowitz@cita.utoronto.ca}

\altaffiltext{2}{Harvard-Smithsonian Center for Astrophysics, 60
Garden Street, Cambridge, MA, 02138, USA; mholman@cfa.harvard.edu}

\begin{abstract}
We examine the eccentricity evolution of a system of two planets
locked in a mean motion resonance, in which the outer planet loses
energy and angular momentum. The sink of energy and angular momentum
could be either a gas or planetesimal disk. We show that the
eccentricity of both planetary bodies can grow to large values,
particularly if the inner body does not directly exchange energy or
angular momentum with the disk. We analytically calculate the
eccentricity damping rate in the case of a single planet migrating
through a planetesimal disk. We present the results of numerical
integrations of two resonant planets showing rapid growth of
eccentricity. We also present integrations in which a Jupiter-mass
planet is forced to migrate inward through a system of $5-10$ roughly
Earth mass planets. The migrating planet can eject or accrete the
smaller bodies; roughly $5\%$ of the mass (averaged over all the
integrations) accretes onto the central star. The results are
discussed in the context of the currently known extrasolar planetary
systems.
\end{abstract}

\keywords{planetary systems---stars: }

\section{INTRODUCTION}
The sixty or so extrasolar planetary systems known to date have
revealed three striking features (for an up to date list of systems
and their properties see http://www/exoplanets.org/ or
http://www.obspm.fr/encycl/encycl.html). First, the distribution of
orbital semimajor axes of the planets range from $\sim3$ AU down to an
almost incredible $0.038$ AU. Second, most of the objects have high
eccentricity by solar system standards, with a typical value being
around $e=0.4$, but ranging up to $0.927$. Third, the parent stars are
highly metal rich, and appear to have accreted iron rich
material after having reached the main sequence
\cite{gonzalez,santos,laughlin,murray}.

The simplest interpretation of the small orbits is that Jupiter-mass
planets experience large scale migrations in some cases, but not in
others; Jupiter falls into the latter class. There are currently two
viable explanations for the migration, tidal interactions between the
planet and the gas disk out of which it formed \cite{GT,lbr}, and gravitational
interactions between the planet and a massive (1-5 Jupiter mass)
planetesimal disk (Murray et al 1998).

The most straightforward interpretation of the high eccentricities,
that they result from collisions or near collisions of two or more
Jupiter-mass planets, is appealing, but require that most systems are
dynamically unstable, in addition to undergoing
migration. Furthermore, a recent exhaustive study of the problem
indicates that the number of systems with low eccentricities is
smaller than would be produced by collisions and scattering \cite{ford}.

In this paper we investigate another possible mechanism for producing
large eccentricities; resonant migration. We suppose that a
Jupiter-mass planet is forced to migrate inward, either by tidal
torques or by ejection of planetesimals, and that a second (possibly
much less massive object) is in a mean motion resonance with the
first. We further assume that the migration process does not
significantly damp the eccentricity of the inner body. This could
occur in migration in a gas disk if the inner disk manages to drain
onto the central star while leaving behind the planets and a
substantial outer gas disk. It would almost inevitably occur in
migration through a massive planetesimal disk, since the planetesimals
are likely to accrete into terrestrial mass or larger bodies; we show
below by direct numerical integrations that these $1-50M_\oplus$
bodies will be trapped into mean motion resonances.

We show that the inward migration of two planets trapped in a mean
motion resonance can produce eccentricities as high as $0.7$. We also
show that in the case of migration by planetesimal ejection, that the
final state may or may not have two resonant planets. Whether the
distribution of eccentricity with planetary mass and semimajor axis
produced by such resonant migrations is consistent with the observed
distribution is a question left for later work.

As a byproduct of our numerical simulations, we find that the fraction
of planetesimal disk mass that accretes onto the star is likely to be much
smaller than found in the work of \cite{qh}; that work studied
the accretion of massless test particles subject to gravitational
perturbations from a migrating Jovian-mass planet. The authors found
that of order half the mass in the disk would accrete. Using our more
realistic, but less extensive integrations of massive planetesimals,
we find a much smaller fraction ($\sim5\% $) of the disk mass
accreting onto the star in the early stages of the migration. (Another
$\sim5-10\%$ of the disk mass will fall on the star if the planet
approaches within $\sim0.1$ AU)\cite{hansen}.

The remainder of the paper is organized as follows. In contrast to
tidal torque migration the eccentricity evolution of a Jupiter-mass
object migrating through a massive planetesimal disk has not been
extensively studied. Section 2 gives a short derivation in the case
that the migrating planet does not accrete a substantial fraction of the
planetesimals. This is appropriate when the escape velocity from the
surface of the planet is larger than the escape velocity at the
orbital distance of the planet from the star. Section 2 describes the
process of capture into resonance, and the evolution of the
eccentricities of both resonant bodies as the migration
proceeds. Section 4 presents the results of numerical integrations of
two resonant bodies, with parameters appropriate for planetesimal
migrations, as well as integrations involving up to $11$
planets. Section 5 gives a discussion of our results, and contrasts
the two types of migration, while section 6 presents our conclusion.

\section{MIGRATION AND ECCENTRICITY EVOLUTION}
We examine the eccentricity evolution of a
Jupiter-mass body migrating inwards due to the extraction of energy
and angular momentum.  The energy $\Ep$ and angular
momentum $\Lp$ of the planet are given by
\begin{eqnarray} 
&\Lp=m_p\sqrt{GM_*a_p(1-\ep^2)}\label{ELL}\label{energy}\\
& \Ep=-{GM_*m_p\over 2a_p}\label{ELE} 
\end{eqnarray} 
Taking the time derivative of equation (\ref{energy}), we find the
time variation of $\ep$ in terms of the time variation of the
planetary energy, assuming the planetary mass is
fixed:
\be \label{ecc_evol}
{\ep\over 1-\ep^2}{d\ep\over dt}
=-{1\over2}\left({1\over\Ep}{d\Ep\over dt}\right)
\left[1+2\left({\Ep\over \Lp}{d\Lp\over d\Ep}\right)\right].
\ee 
The quantity 
\be \label{beta}
\beta\equiv\left[1+2\left({\Ep\over \Lp}{d\Lp\over d\Ep}\right)\right]
\ee 
is a convenient measure of the rate at which the planetary
eccentricity changes.  Both $E_p$ and $dE_p/dt$ are negative for an
inward migration, so $e_p$ decreases if $\beta>0$. Conservation of
energy and angular momentum implies that $d\Lp/d\Ep=(dL/dE)_T$, where
the latter quantity is the ratio of the rates at which angular
momentum and energy are removed from the system by whatever process is
driving the migration.  The planetary eccentricity decreases as long
as
\be \label{damping}
\left({dL\over dE}\right)_T<-{\Lp\over2\Ep}=\sqrt{1-\ep^2}\Big/n_p.
\ee 

We now specialize to the case of planetesimal migration.  To find
$dL_p/dE_p$, we calculate the total change in $E$ and $L$ for a
planetesimal of mass $m$ (where $m<<m_p$) from its initial orbit, with
semimajor axis $a$ and eccentricity $e$, to the point at which it is
ejected, then use conservation of energy and angular momentum. We note
that this is not adequate for cases where the planet eats the
planetesimal, since some orbital energy will be lost in the form of
radiation in that case.

\begin{figure}
\plotone{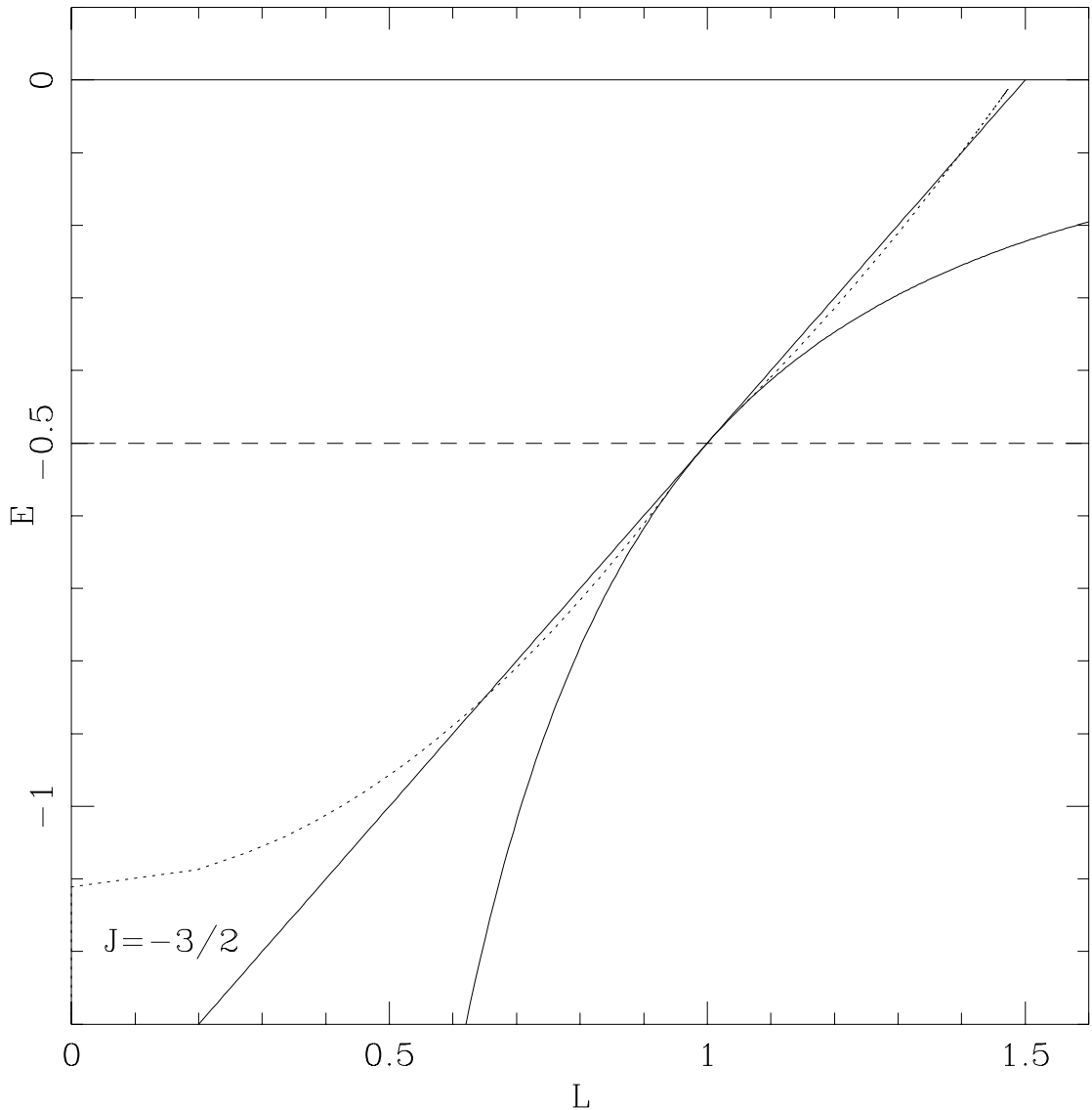}
\caption[The $E-L$ plane]{The $E-L$ plane for a planet with
$e_p=0.1$. The curved solid line is the maximum angular momentum
possible for a body of the given energy; it corresponds to a circular
orbit ($e=0$). The dotted line corresponds to a planetesimal orbit that just
grazes the orbit of the Jupiter mass body. The solid line labeled
$J=-3/2$ illustrates the minimum value of the Jacoby parameter that
must be reached by the asteroid in order to be lifted from an orbit
with $a<a_p$ to an orbit with $a>a_p$. In this figure we employ units
in which $GM_*=\ap=1$.
\label{Fig_EL}}
\end{figure}

Figure (\ref{Fig_EL}) illustrates the constraints on the evolution of
a planetesimal in the $E-L$ plane, in the case $e_p=0.1$. The solid
curved line corresponds to $e=0$; planetesimal orbits must lie to the
left of this line. In analyzing motion involving close encounters
with a planet, it is useful to introduce the Jacoby parameter,
$J=E-n_pL$, where $n_p$ is the mean motion of the planet. The diagonal
solid line corresponds to $J/m=-3/2$ (in units where $GM_*=\ap=1$). We
note that in order to be ejected from $a<\ap$ the asteroid must reach
$J/m\ge -3/2$, since it must pass through $a=a_p$ with $e\ge0$ to be
ejected.

We will assume that the planetesimal is ejected with zero total
energy; if it is ejected with a larger energy, the damping rate will
be smaller than the estimate we obtain below. The initial energy and
momentum of the planetesimal are given by expressions analogous to
equations (\ref{ELL}) and (\ref{ELE}).  Rather than calculating the
change in $L$ directly, we calculate the change in the Jacoby
parameter; we do so because $J$ is constant (in a
statistical sense) during the planet crossing phase of the asteroid's
evolution \cite{opik}. To lowest order in $m/m_p$ we have
\be \label{dlde}
\left({dL\over dE}\right)_T=\left[1-{dJ\over dE}\right]\Bigg/n_p.
\ee 
If the planetesimal disk is originally cold ($e<<1$) and $e_p<<1$, few
planetesimals will cross the orbit of the planet. However,
planetesimals trapped in resonance with the planet will suffer chaotic
perturbations which on average transfer angular momentum, but not
energy, from the asteroid to the planet. This causes $J$ to increase
while leaving $E$ fixed. Once enough angular momentum has been removed
from the asteroid's orbit, the asteroid can suffer close encounters
with the planet. We assume that the first close encounter removes the
asteroid from the resonance, while leaving $J$ fixed. Subsequent
encounters extract or supply energy and angular momentum to the
asteroid in such a way as to leave $J$ constant on average, as noted
above. Eventually the planetesimal is ejected with $E\ge0$; taking
$E=0$ we find
\be \label{djde1}
{dJ\over dE}=2\left({a\over \ap}\right)^{3/2}
\left(\sqrt{1-e^2}-\sqrt{1-e_c^2}\right),
\ee 
where $e_c\equiv \ap(1-\ep)/a-1$ is the eccentricity at which the
planetesimal just crosses the orbit of the planet. Note that $dJ/dE\ge0$.

In arriving at equation (\ref{djde1}) we have assumed that the final
Jacoby parameter 
\be 
J_f/m=J_c/m\equiv-{1\over 2a}-\sqrt{a(1-e_c^2)}>-3/2
\ee 
(we again use $GM_*=\ap=1$). If the Jacoby parameter at planet
crossing ($J_c$) is not larger than $-3m/2$, the planetesimal must
diffuse to higher $J$ in order to be ejected, since it has to get past the
solid curve in Figure (\ref{Fig_EL}). Setting $J_f/m=-1.5$ we find
\be \label{djde2}
{dJ\over dE}=2\left({a\over \ap}\right)^{3/2}
\left[\sqrt{1-e^2}+{1\over 2}\left({\ap\over a}\right)^{3/2}-{3\over2}
\left({\ap\over a}\right)^{1/2}\right].
\ee 
When $J_c/m>-3/2$ equation (\ref{djde1}) should be used, while equation
(\ref{djde2}) is appropriate if $J_c/m<-3/2$.

Combining equations (\ref{ecc_evol}) and (\ref{dlde}), the expression
for the rate of change of the planet's eccentricity is
\be \label{planetesimal_damping}
{\ep\over1-\ep^2}{d\ep\over dt}=-{1\over2}
\left({1\over\Ep}{d\Ep\over dt}\right)
\left[1-{\left(1-dJ/dE\right)\over \sqrt{1-\ep^2}}\right]
\ee 

\begin{figure}
\plottwo{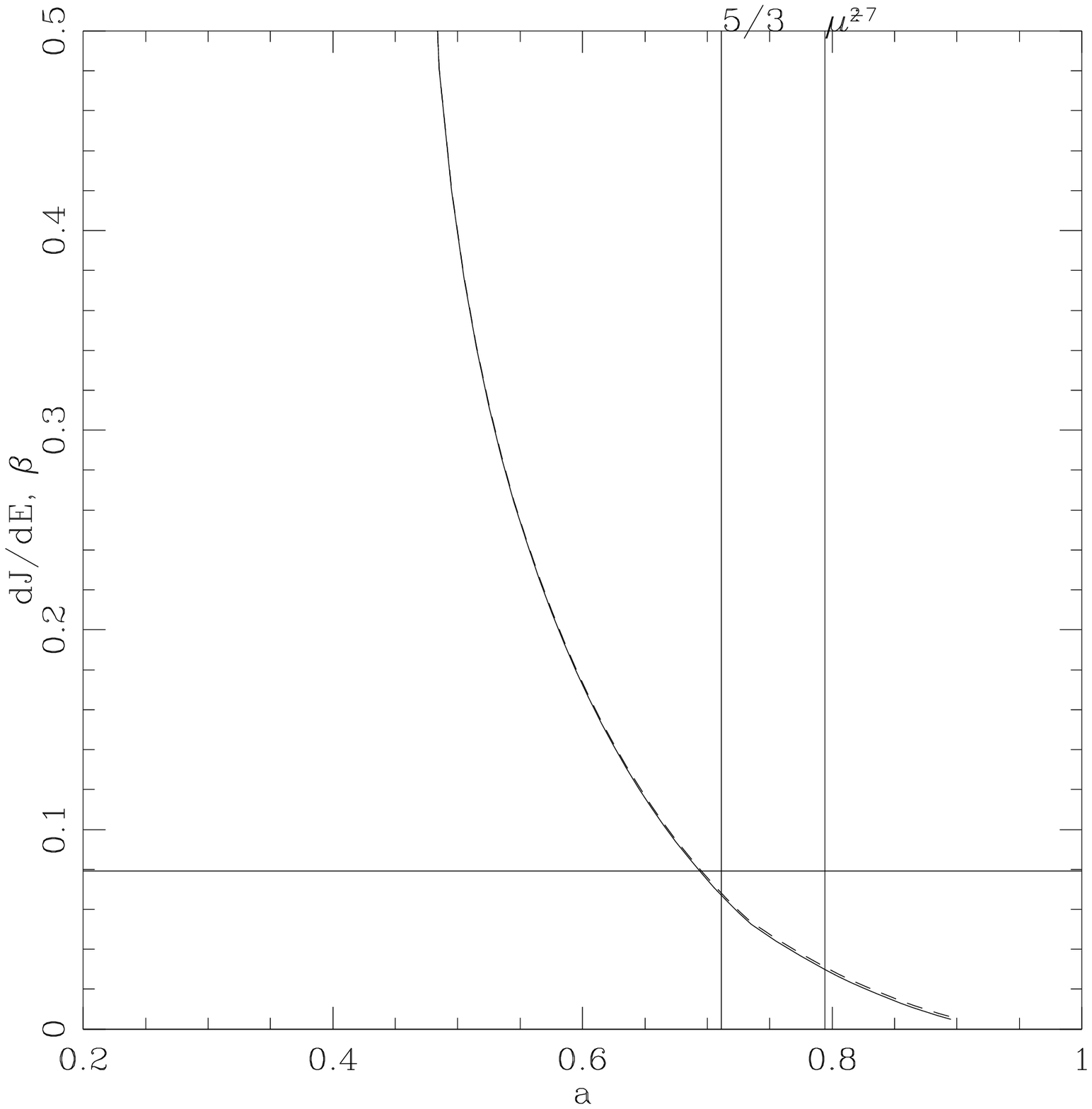}{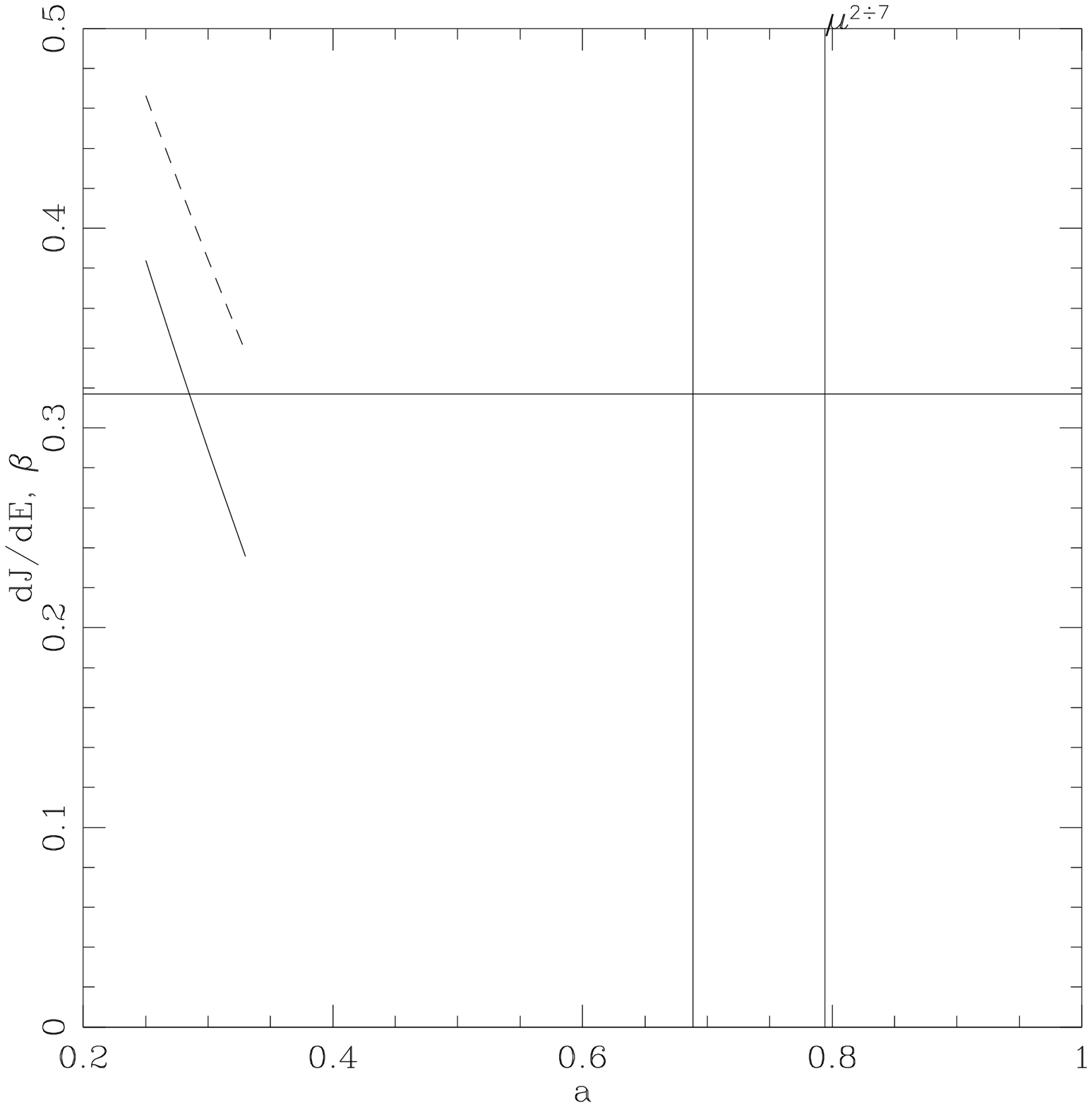}
\caption[$dJ/dE$]{a) The time averaged change $dJ/dE$ for a
planetesimal with initial eccentricity $e=0.05$, as a function of its
initial $a$ (dashed curve), and the corresponding $\beta$ (solid
curve). The planet is assumed to have a mass equal to that of Jupiter
($318M_\oplus$), and $e_p=0.05$. The solid vertical line marked $\mu^{2/7}$
marks the region where first order mean motion resonances overlap,
producing large scale chaos. The solid vertical line near $a=0.71$
marks the location of the $5/3$ mean motion resonance. Note that
$dJ/dE\approx 0.03$ near the $\mu^{2/7}$ region. The horizontal line
corresponds to $\gamma$ for an inner planet of mass $10M_\oplus$
trapped in the $4/1$ mean motion resonance.  b)
Same as in (a), except that $e_p=e=0.5$. The horizontal line
corresponds to an inner planet with a mass of $40M_\oplus$.
\label{Fig_dJdE}}
\end{figure}

It can be shown, using equations (\ref{djde1}) and (\ref{djde2}), that
\be 
{dJ\over dE}\ge1-\sqrt{1-\ep^2}.
\ee 
In other words, planetesimal migration, as described here, always
damps the eccentricity of a single planet. We can estimate the value
of $dJ/dE$ when $e$ and $\ep$ are small; for example for $J_c<-3/2$
(which requires $e\lta (3+2\sqrt{3})\ep$) equation
(\ref{djde2}) evaluated at the maximum $a/\ap=(1-\ep)/(1+e)$ gives
\be 
{dJ\over dE}\approx {3\over4}\ep^2+{3\over 2}\ep e-{1\over 4}e^2.
\ee 
For $e\approx \ep=0.05$, $dJ/dE\approx 2\ep^2\approx0.005$. For
smaller values of $a$ this increases, as shown in Figure
\ref{Fig_dJdE}. Panel (a) in the Figure shows $\beta$ and $dJ/dE$ for
a planet with $\ep=0.05$ ejecting a planetesimal with initial
$e=0.05$, as a function of the initial $a$ of the planetesimal. The
relevant value of $dJ/dE$ depends on the average initial $a$ of the
planetesimals that are ejected, i.e., it depends on which resonance is
most actively ejecting objects. Early on in the evolution of the
system we expect the planetesimal disk to be truncated inside the
chaotic zone produced by the overlap of first order mean motion
resonances (the ``$\mu^{2/7}$ chaotic region''\cite{wisdom}). Under
those circumstances the relevant resonance is the $5/3$, at
$a/\ap\approx 0.71$; $dJ/dE\approx 0.066$ for that case. However, as
the migration proceeds, material is supplied to the $\mu^{2/7}$ zone,
and $dJ/dE$ will be on average smaller, at least while $e_p<<1$. As
$e_p$ increases, equations (\ref{djde1}) and (\ref{djde2}) show that
$dJ/dE$ no longer increases as rapidly as $e_p^2$; it effectively
saturates near $dJ/dE\approx0.3-0.4$. Panel (b) shows $\beta$ and
$dJ/dE$ in the case $\ep=e=0.5$. Planetesimals never get the chance to reach
the $5/3$ resonance, since they become planet crossing at much smaller
semimajor axis. The damping rate is of order $0.35$. 

We can calculate the circularization time for planetesimal migration,
in terms of the migration time. Following the notation employed in
satellite studies \cite{lpc},
\begin{eqnarray} 
{d\Ep\over dt}&=&-(\np T+H)\label{torque}\\
{d\Lp\over dt}&=&-T,
\end{eqnarray} 
where $T$ is the (average) torque exerted on the planet by the
ejection of planetesimals, and $H$ is responsible for removing energy
from the radial motion of the planet, i.e., it damps the eccentricity.
(Actually it would be better to use
$d\Ep/dt=-(\np T/\sqrt{1-\ep^2}+H)$, as will become apparent). It
follows from these equations that
\be 
H=-\beta{d\Ep\over dt},
\ee 
where we have replaced the missing factor of
$\sqrt{1-\ep^2}$. Using this in equation (\ref{torque})
we define the migration time 
\be 
\tau_m\equiv {GM_*\mp\over 2\ap\np T}(1-\beta)
\ee 

The circularization time $\tau_c$ is given by equation (\ref{ecc_evol});
\be 
\tau_c\equiv \tau_m{\ep^2\over \beta(1-\ep^2)}.
\ee 
For small $\ep$ this reduces to $\tau_c\approx(2/3)\tau_m$; the
circularization time is comparable to the migration time.

Note that in deriving equations (\ref{djde1}) and (\ref{djde2}) we
have ignored the finite extent $D\equiv(m_p/3M_*)^{1/3}a_p$ of the
planet's Hill sphere, the region over which the planet's gravity
exceeds the tidal acceleration from the central star. For a Jupiter
mass planet $D\approx 0.07a_p$. Including the effect of the Hill
sphere in our analysis effectively increase $\ep$ to
$\ep'=\ep+(\mp/3M_*)^{1/3}$; this is why we use $\ep=0.1$ in Figure
\ref{Fig_EL}.

\section{RESONANCE CAPTURE AND ECCENTRICITY EVOLUTION}
We have seen that a single planet embedded in a planetesimal disk
suffers eccentricity damping (it appears that a similar statement
applies to a single planet in a gas disk \cite{PNM}). However, a
Jupiter mass planet migrating through a disk of planetesimals will
capture bodies into resonance; in the early stages this is how the
migration proceeds. We show in this section that these resonant bodies
tend to increase the eccentricity of the Jupiter mass planet; if
$10-20$ Earth masses (denoted $M_\oplus$) are trapped into a
resonance, then this resonant eccentricity driving exceeds the
eccentricity damping described in the previous section, and the
eccentricity of the planet will increase as it migrates inward.

We begin by describing capture into resonance. We consider the
gravitational interaction of two planets in orbit around a much more
massive central body. For simplicity we consider only the planar
problem.  In the absence of dissipative effects the Hamiltonian
describing the motion is
\be  \label{Hamiltonian}
H=-{\mu_1^2m_1\over 2L_1^2}-{\mu_p^2\mp\over 2\Lp^2}-
{Gm_1\mp\over a_1}\sum_{\bf j}\Phi_{\bf j}(a_1,\ap)e_1^{|j_3|}\ep^{|j_4|}
\cos\left[j_1\lambda_1-\jp\lambda_p+j_3\varpi_1+j_4\varpi_p
\right].
\ee  
Here $\mu_1\equiv{\cal G}(\ms+m_1)$, where ${\cal G}$ is Newton's
gravitational constant, and $L_1=\sqrt{\mu_1 a_1}$, with similar
definitions for the outer planet (labeled with a subscript $p$).  The
third term in equation (\ref{Hamiltonian}) represents the mutual
perturbations of the two planets. It produces variations in the
orbital elements ($a$, $e$, and so forth) of order the planetary mass
$m_i$.

The coefficient $\Phi\sim \left[a_1/(\ap-a_1)\right]^{|j_1-\jp|}$
\cite{hm}. The integers $j_i$ satisfy the relation
$j_1-\jp+j_3+j_4=0$. Each cosine term in the sum is referred to as a
resonant term or simply as a resonance. Resonances with $|j_1-\jp|=q$
are proportional to $q$ powers of eccentricity, and are said to be
$q$th order mean motion resonances.  The planets are said to be in
resonance if one or more of the arguments of the cosines are
bounded. Since $n_1\equiv\left<\dot\lambda_1\right>$ (where the angle
brackets refer to an average over a single orbit) and $\np$ are much
larger than $\dot\varpi_1$ and $\dot\varpi_p$, the condition for resonance
is roughly equivalent to
\be \label{lock}
j_1n_1-\jp\np=0, 
\ee 
or $\ap/a_1=(\jp/j_1)^{3/2}$. Throughout this section we ignore
non-resonant terms of second order in the planetary masses.

Suppose that $\ap/a_1$ is initially larger than this resonant value,
but that some dissipative process acts to reduce $\ap$ while leaving
$a_1$ unchanged. Then the torques represented by the resonant cosine
term will increase, since $\Delta a\equiv \ap-a_1$ is decreasing and
the torques are proportional to $(a_1/\Delta a)^2$. Another way to say
this is that the depth of the potential well represented by the
resonant term is increasing, as is the width of the resonance. As
$\ap/a_1$ passes through the resonant value $(j_1/\jp)^{2/3}$, the
planets may be trapped into resonance. The torques represented by the
resonant term in eqn. (\ref{Hamiltonian}) will then transfer energy
and angular momentum between the two planets in just such a way as to
maintain the resonance while both bodies move toward the star
\cite{goldreich}.

Capture is much less likely if the planets are moving away from each
other, for example if $a_1$ is decreasing, or if both semimajor axes
are decreasing but that of the inner planet decreases more rapidly; in
that case the size of the resonance is decreasing, and the planets
will usually pass through the resonance without being trapped.

Henceforth we will assume that the outer planet is moving toward the
inner planet, resulting in capture into resonance. 

Suppose for the moment that the inner planet has a sufficiently small
mass that it cannot effectively scatter planetesimals, so that it does
not lose energy or angular momentum directly to the planetesimal
bath. (In the case of gas migration, we assume that the inner disk is
non-existent). By virtue of its resonance interaction with the outer planet, it
nevertheless does supply energy and angular momentum indirectly to
material driving the migration. Another way to
say this is that $dE_p/dL_p$ no longer equals $(dE/dL)_T$, the quantity
calculated for the case of planetesimal migration in the previous section.
Here we calculate the relation between these two quantities that
obtains when a second object of mass $m_1$, is in
resonance with the Jupiter-mass object. 

We relate $dE_p/dL_p$ to $(dE/dL)_T$ using the conservation of energy
and angular momentum. Conservation of energy gives
\be 
{dE\over d\Ep}=1+{m_1\over \mp}\left({\jp\over j_1}\right)^{2/3},
\ee 
while conservation of angular momentum implies
\be 
{d\Lp\over d\Ep}=\left({dL\over
dE}\right)_T\left[\dEdEp\right]-{dL_1\over dE_p}
\ee 
We also need
\be 
{L_1\over\Lp}=\LLp.
\ee 
Using all these in (\ref{ecc_evol}) we find
\be 
{\ep\over1-\ep^2}{d\ep\over dt}=-{1\over2}
\left({1\over\Ep}{d\Ep\over dt}\right)
\left[
1+2{\Ep\over\Lp}\left( {dL\over dE} \right)_T \left( {dE\over d\Ep}\right)
\right] 
+{L_1\over\Lp}\left({1\over L_1}{dL_1\over dt}\right).
\ee 
Note that the $dE/d\Ep$ factor multiplying $(dL/dE)_T$ is larger than
one; it effectively increases $(dL/dE)_T$. From equation
(\ref{damping}) we see that this will tend to increase $\ep$. The term
proportional to $dL_1/dt$ will tend to decrease $\ep$, but we shall
see that its effect is smaller than that of the term involving
$dE/d\Ep$; this is the origin of the increase in eccentricity of two
resonant bodies undergoing migration.  From the expression for $L_1$,
and using the resonance condition, we have
\be 
\left({1\over L_1}{dL_1\over dt}\right)=
-{1\over 2}\left({1\over\Ep}{d\Ep\over dt}\right)
-{e_1\over 1-e_1^2}{de_1\over dt}.
\ee 
As just noted the first term on the right will tend to damp the
eccentricity of the outer planet; the second term on the right will
also damp $\ep$ as long as $de_1/dt>0$. Combining the last two
equations we find
%
%
%
\begin{eqnarray} \label{almost_there}
{\ep\over1-\ep^2}{d\ep\over dt}&=&-{1\over2}
\left({1\over\Ep}{d\Ep\over dt}\right)
\Bigg\{
\left[1+2{\Ep\over\Lp}\left( {dL\over dE} \right)_T\right]
\left( {dE\over d\Ep}\right) \nonumber\\
&&-{m_1\over\mp}\left({j_1\over \jp}\right)^{1/3}{1\over\sqrt{1-\ep^2}}
\left[
\left({\jp\over j_1}\right)\sqrt{1-\ep^2}-\sqrt{1-e_1^2}
\right]
\Bigg\} \nonumber\\
&&-{m_1\over\mp}\left({j_1\over \jp}\right)^{1/3}\sqrt{1-e_1^2\over
1-\ep^2}
{e_1\over 1-e_1^2}{de_1\over dt},
\end{eqnarray} 
where we have added and subtracted $(m_1/\mp)(\jp/j_1)^{2/3}$ inside
the curly brackets.

We identify 
\be 
\left({d\Ep\over dt}\right)_T\equiv\left({d\Ep\over dt}\right)\left( {dE\over d\Ep}\right)
\ee 
as the rate at which the expulsion of planetesimals removes energy
from the outer planet. With this identification, and recalling
equation (\ref{ecc_evol}), it becomes clear that
the first term in curly brackets in equation (\ref{almost_there}) is
the time rate of change of $\ep$ due to the expulsion of
planetesimals. The final result is
\begin{eqnarray} 
{\ep\over1-\ep^2}{d\ep\over dt}&=& 
-{1\over2}\left({1\over \Ep}{d\Ep\over dt}\right)_T
\Bigg\{
1-{(1-dJ/dE)\over \sqrt{1-\ep^2}}\nonumber\\
&&-{m_1\over\mp}\left({j_1\over \jp}\right)^{1/3}{1\over\sqrt{1-\ep^2}}
\left[
\left({\jp\over j_1}\right)\sqrt{1-\ep^2}-\sqrt{1-e_1^2}
\right]\Bigg/\left[\dEdEp\right]
\Bigg\} \nonumber\\
&&-\LLp{e_1\over1-e_1^2}{de_1\over dt}\label{ep}
\end{eqnarray} 
Note that this expression reduces to equation
(\ref{planetesimal_damping}) when $m_1\to0$.

Equation (\ref{ep}) has two undetermined quantities ($d\ep/dt$ and
$de_1/dt$). Bodies that are trapped in a mean motion resonance typically have
their apsidal lines locked as well, so that
$\dot\varpi_1=\dot\varpi_2$. Using the equations of motion for
$\varpi_1$ and $\varpi_2$, we can find a relation between $e_1$ and
$\ep$ that depends on the precession rates of the apsidal lines. The
latter are determined both by the mutual perturbations of the two
planets, and by the distribution of mass in the planetesimal (or gas)
disk. Given the current state of both observations and theory, we feel
that a detailed calculation is not justified.

In the appendix we present another derivation in which we allow for
the possibility that the inner, less massive planet also loses energy
and angular momentum to the sink of energy and angular momentum.

We proceed to examine some limiting cases. First, suppose that no
tides act on the inner planet, and that the tides acting on the
outer planet keep that planet's orbit circular. Then
\be 
{e_1\over \sqrt{1-e_1^2}}{de_1\over dt}\approx
-\left({d\ln \ap\over dt}\right)_T 
\left[
{\jp\over j_1}-\sqrt{1-e_1^2}
\right]\Big/2\left(dE/d\Ep\right).
\ee 
Since $\ap$ is decreasing, the right hand side is positive, and $e_1$
will grow. Assuming $\jp/j_1>>1$, we find
\be \label{e_1_of_a}
e_1\approx\sqrt{1-(1-\gamma_1\ln a_{1,i}/a_1)},
\ee 
where $a_{1,i}$ is the semimajor axis of the inner body when it is
captured into resonance,  $\gamma_1=\jp/2j_1$, and we have neglected
the initial value of $e_1$.

Now suppose that no tides act on the inner body, and that the
migration has proceeded far enough that $e_1$ has grown to the point
that $de_1/dt$ is small. Then we can find the equilibrium value for
$\ep$ by setting the terms in the curly brackets equal, assuming that
$\beta=2\ep^2/3$;
\be \label{emax}
e_{p,max}\approx\sqrt{2m_1\ap\over 3(\mp a_1+m_1\ap)},
\ee 
where we neglect $\sqrt{(1-e_1^2)/(1-\ep^2)}$ compared to
$\jp/j_1$. For gas-disk migration, the ratio of $\tau_c/\tau_m$ would
enter in the expression for $\beta$. For two equal mass bodies in a
$4/1$ resonance this is about $0.7$. Even for $m_1/\mp=0.1$ (a
$30M_\oplus$ inner planet) the equilibrium eccentricity is
$0.41$. However, note that this expression is actually an
underestimate in the case of planetesimal migration, since $\beta$,
which is a measure of the damping due to the migration process, does
not scale as $\ep^2$ for $\ep$ as large as $0.4$; $\beta$ actually
grows less rapidly, meaning that the damping is not as efficient as
equation (\ref{emax}) assumes.

We can relate the eccentricity to the distance migrated when
$\mp>>m_1$. Define
\be 
\gamma\equiv{m_1\over \mp}\left({\jp\over
j_1}\right)^{2/3}.
\ee 

The evolution of the eccentricity is then described by
\be \label{evolution}
{1\over 1-\ep^2}{d\ep^2\over dt}\approx 
-{1\over2}\left({1\over\Ep}{d\Ep\over dt}\right)_T
\left[\beta-\gamma\right]
\ee 
We have assumed that $\jp/j_1>>\sqrt{1-e_1^2}$, and neglected terms
second order in $m_1/\mp$. The eccentricity of the outer planet will
grow indefinitely providing $ \gamma>\beta$.

Now suppose that $\beta$ is independent of $\ep$; this is not true
when $\ep$ is small, but for small $\ep$ $\gamma$ can be much larger
than $\beta$; for $\ep\approx0.2$ or larger $\beta$ is
roughly constant. In that case we can integrate equation
(\ref{evolution}) to find the final eccentricity $e_{2f}$ of the outer
body,
\be \label{e_2_of_a}
e_{2f}\approx\sqrt{1-\left({a_{2f}\over a_{2i}}\right)^{\gamma-\beta}},
\ee 
where $a_{2i}$ is the semimajor axis and eccentricity of
the outer planet when it enters the resonance, and $a_{2f}$ is the
semimajor axis when the planet either stops migrating or leaves the
resonance. We have assumed that the initial eccentricity $\ep<<1$; if it
is not, the final eccentricity will be larger.

The numerical work described below shows that a Jupiter-mass object
can capture smaller bodies into resonances ranging from the $2/1$ to
the $4/1$; we have even seen captures into the $11/2$ resonance. In a
scenario where the Jupiter-mass body migrates through a planetesimal
disk having a comparable mass, we expect resonance capture of
terrestrial bodies with masses ranging from 1 Earth mass ($M_\oplus$)
up to $30M_\oplus$ or more; an extreme upper limit might be of order
$50M_\oplus$, corresponding to about $10\%$ of the disk mass. The
plausible range for $\gamma$ is then $5\times10^{-3}-0.4$.  For
migration in a planetesimal disk we expect $\beta$ to be in the range
$0.01$ (for $\ep<<1$) to $0.3$ (for $\ep\gta0.5$).  Taking
$1$ AU as a representative value for $a_{2f}$ (although some
extrasolar planets are in much smaller orbits), with $a_{1f}\sim5-10$
AU, we find final eccentricities in the range $0.1-0.6$, with
$\ep\approx0.45$ being a typical value.

\section{Numerical Results}
We employ a Bulirsch-Stoer integrator with a variable step size. We
require that at each time step the relative accuracy of the
integration (as measured in phase space) be $10^{-12}$. Typical
orbital times are of order one to ten years, while the integrations
can extend up to $10^8$ yrs. In test runs where no energy is removed
from the planet, the largest variation in total energy is typically
less than a part in $10^9$. In most of the runs reported on here we
remove energy and angular momentum from the largest planet; the
variations in energy and angular momentum from the expected amounts
are similarly small.

In runs with multiple massive planets collisions often occur. We
assume that the smaller planets are rocky bodies with bulk densities
of $3\ {\rm g/cm^3}$, that collision occur when planets are within two
times the sum of the planetary radii (to allow for capture by tidal
disruptions) and that the captures are completely inelastic with no
loss of mass to small fragments. This is reasonable for collisions
involving the more massive objects.

We assume that the most massive body (``Jupiter'') migrates inward by
ejecting numerous planetesimals from the system \cite{mhht}. We
simulate this by extracting energy and angular momentum from the orbit
of Jupiter. We implement this numerically as follows.

The energy and angular momentum of the planet are given by
\be 
E={1\over2}\mp (v_x^2+v_y^2)-{GM_*\mp\over r}
\ee 
and 
\be 
L=\mp(xv_y-yv_x)
\ee 
where $v_x$ and $v_y$ are the velocity of the planet (we suppress the
subscript $p$ for ease of reading). Recall that we assume the planet
and planetesimal are coplanar. Taking time derivatives, we invert to find
\begin{eqnarray} \label{drag}
{dv_x\over dt}&=&\left({1\over \Ep}{d\Ep\over dt}\right)
{\Ep\over \mp {\bf r}\cdot{\bf v}}
\left[x+{v_y\over \np}\sqrt{1-\ep^2}(\beta-1)\right]\nonumber\\
{dv_y\over dt}&=&\left({1\over \Ep}{d\Ep\over dt}\right)
{\Ep\over \mp {\bf r}\cdot{\bf v}}
\left[y-{v_x\over\np}\sqrt{1-\ep^2}(\beta-1)\right],
\end{eqnarray} 
where we have used equation (\ref{beta}).
We assume that the close
encounters that lead to changes in the planetesimal's orbital elements
occur on times much shorter than the orbital period, so that we can
take the position of the planet to be fixed.

Using these equations for numerical work is problematic, since the
vector dot product in the denominator vanishes at peri- and
apoapse. We regularize the equations by multiplying the right-hand
side by $2\sin^2f$, where $f$ is the true anomaly.

The terrestrial mass objects in our simulations do not have high enough
escape velocities to efficiently eject smaller bodies, so we do not
force them to migrate.

We have integrated the equations of motion for the two body problem
(the star and a massive planet) modified to account for the drag
imposed by the ejection of planetesimals, as given in equation
(\ref{drag}), regularized as noted above. The eccentricity $e_p$ and
semimajor axis $a_p$ decay as expected. We discuss the behavior of
systems with more than one massive planet in the following subsections.

\subsection{TWO MASSIVE BODIES \label{double}}
\begin{figure}
\plottwo{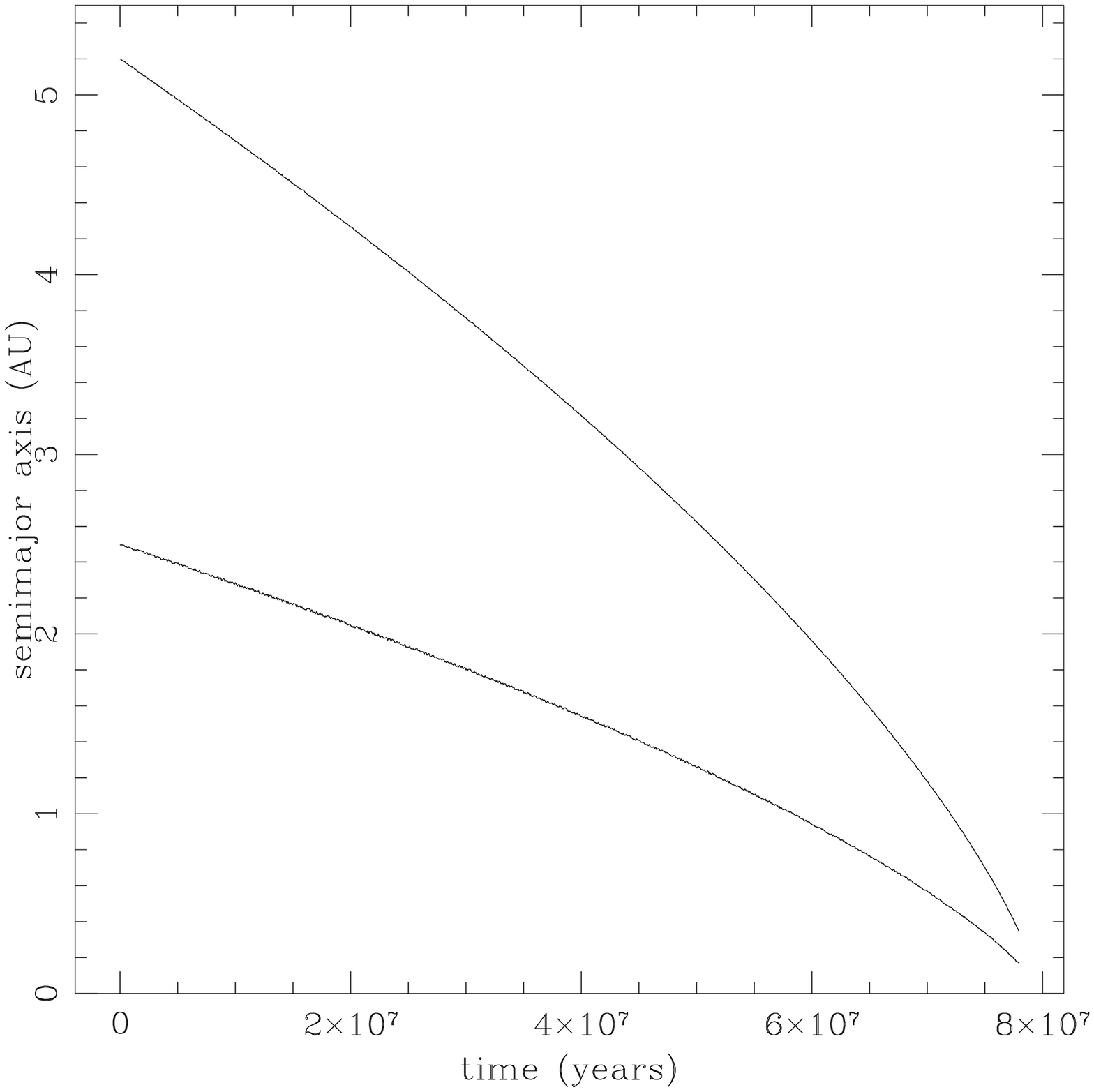}{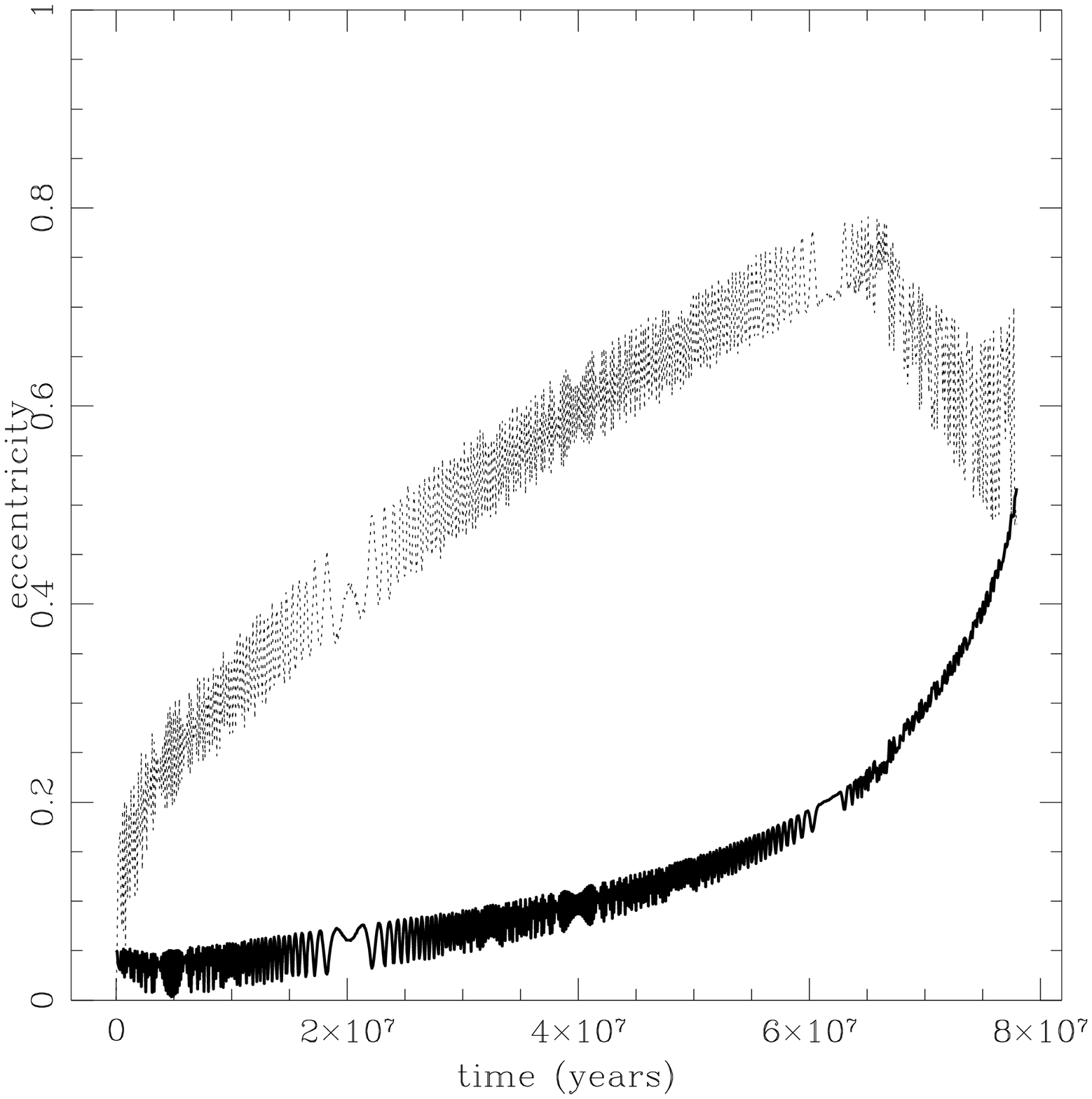}
\caption[$e$ vs. $t$]{The evolution of a system of 2 bodies, one
having a mass of $40M_\oplus$, placed inside the orbit of the second,
a Jupiter mass object that is forced to migrate inward. The inner body
is placed just inside the $3/1$ mean motion resonance. Panel (a) shows
the semimajor axis of both bodies as a function of time, while (b)
shows the eccentricity; the light line (higher $e$) corresponds to the
inner, lighter planet.
\label{Fig_basic}}
\end{figure}

To test the basic idea that migration of two resonant bodies will
induce the growth of eccentricity, we have started two planets just
outside resonance, and applied the ``tides'' described in the previous
paragraphs. An example is shown in Figure (\ref{Fig_basic}). The inner
body has a mass of $20M_\oplus$ and the outer body has a mass equal to
that of Jupiter ($\approx 318M_\oplus$). The upper plot shows the
semimajor axes of both bodies as a function of time, while the lower
plot shows the eccentricities. Energy and angular momentum were
removed only from the outer body. The resonance interaction forces the
inner body to migrate inward as well, as can be seen in the
figure. One can also see from the figure that the eccentricity of both
bodies increased.

The prediction based on (\ref{e_2_of_a}) for the eccentricity of the
outer, more massive body is too large for $\ap$ near the initial
value, which is consistent with the fact that we neglected the rather
rapid variation of $e_1$; as the migration proceeds the prediction
becomes more accurate. By the end of the integration equation
(\ref{e_2_of_a}) is a good approximation.

\subsection{MULTIPLE MASSIVE BODIES\label{multiple}}
The planetesimal disk we postulate is very massive, $1-3$ Jupiter
masses. It is likely that multiply bodies with masses comparable to or
larger than that of the Earth are likely to form in such a massive
disk. As a first step toward a realistic simulation of a migration in
such a disk, we have run a number of cases involving five to ten
roughly Earth mass bodies placed on orbits with random semimajor axes
and small eccentricities inside the orbit of a Jupiter-mass planet. We
then force the Jupiter mass body to migrate inward toward the Earth
mass planets.

\begin{figure}
\plottwo{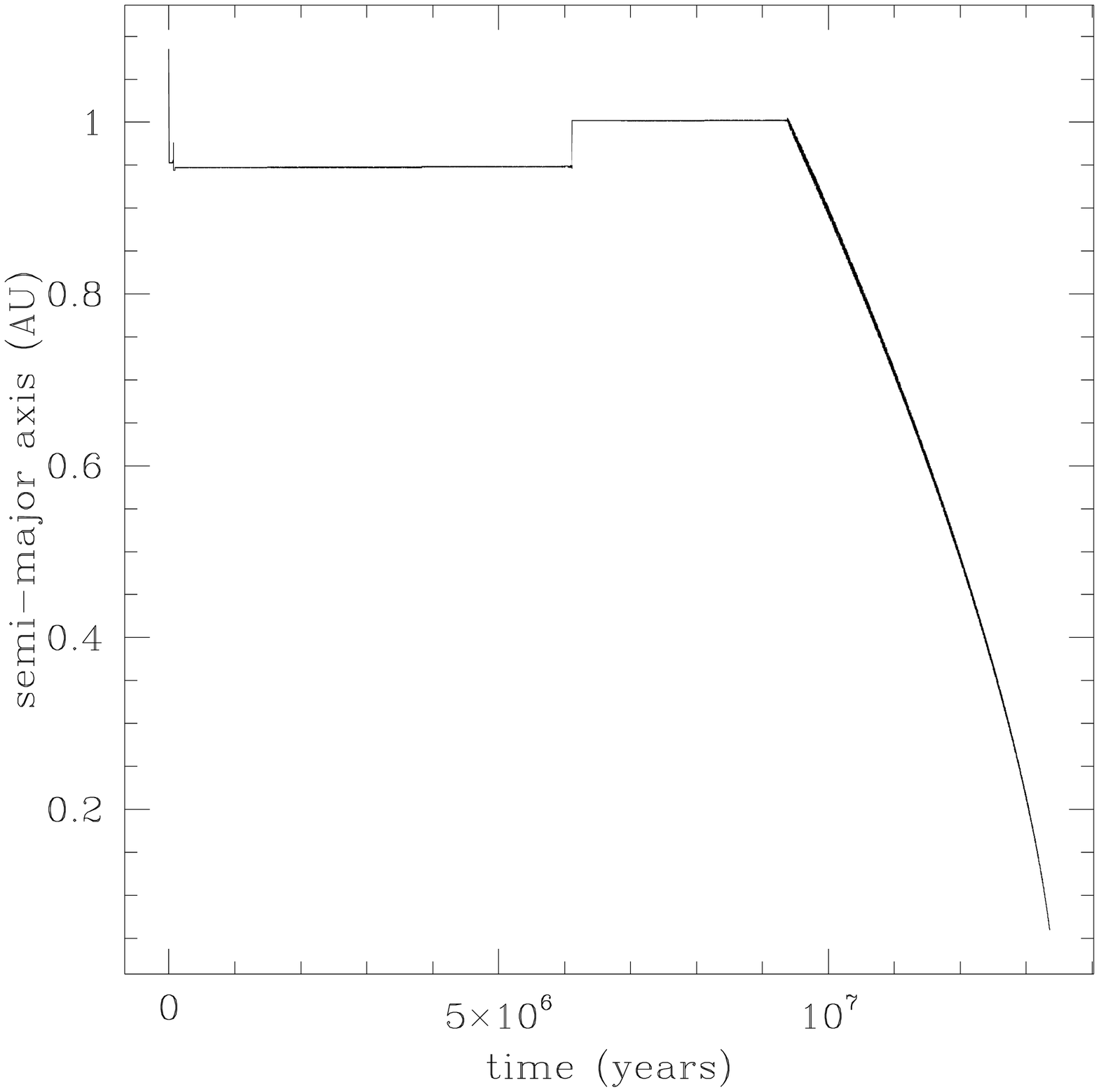}{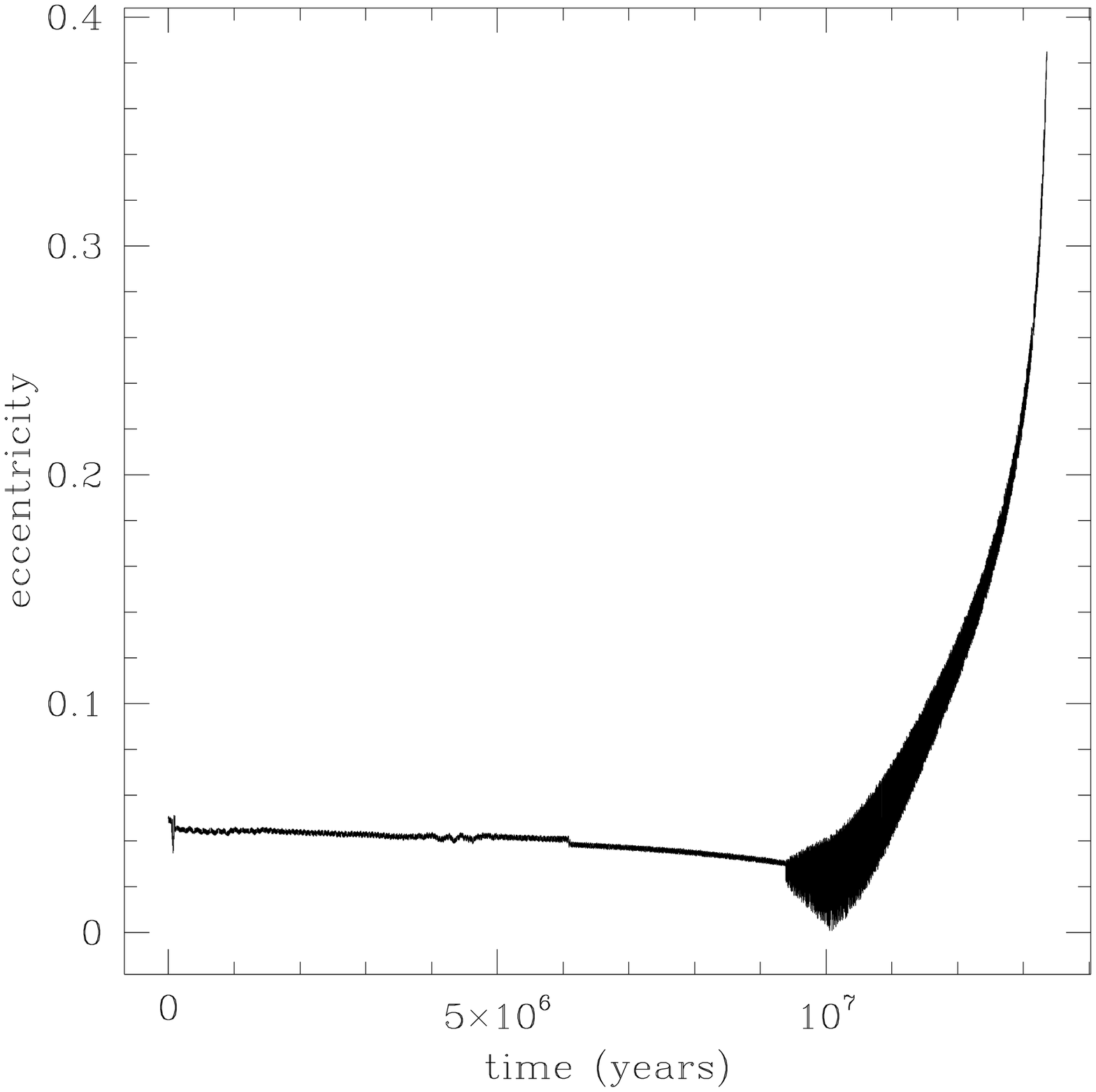}
\caption[crash]{The evolution of a system of 5 bodies with masses
randomly distributed between $0.3$ and $10M_\oplus$, placed inside the
orbit of a Jupiter mass object that is forced to migrate inward. Three
of the small planets merged with each other to create a $6.3M_\oplus$
mass planet. One of the small planets merged with the Jupiter mass
planet, while the fifth small planet was ejected. Panel (a)
shows the semimajor axis of one of the low mass planets that merged to
form the $6.3M_\oplus$ mass body; the mass
increases with time due to collisions with other bodies, seen as jumps
in $a$. This body was caught into resonance with the Jupiter mass
planet at $t=9\times10^6$ years. (b) The eccentricity of the Jupiter
mass body as a function of time. It damps slowly up until the time it
captures the smaller mass body in (a), then rises rapidly. The value
of $\beta$ used in this run assumed that $e=0.05$, which is not
appropriate for the final $\ep\approx 0.4$.
\label{Fig_crash}}
\end{figure}

Figure (\ref{Fig_crash}) shows the result of one such integration. We
started five bodies with masses randomly distributed between $0.3$ and
$10M_\oplus$ with semimajor axes between $0.5$ and $4$ AU. As Jupiter
migrated inward, three of the small planets merged to form a
$6.3M_\oplus$ planet, one small planet crashed into Jupiter, and one
small planet was ejected. Both the latter two events illustrate the
migration mechanism we are postulating. 

After the three small planets merged to form a $6.3M_\oplus$ body, the
resulting planet was captured into the $3/1$ mean motion resonance
with Jupiter. Subsequently the eccentricities of both bodies increased
(we employed a rather low value of $\beta/\ep^2$ for this run). Both
planets migrated inward until the inner planet struck Jupiter, when
Jupiter was at $0.12$ AU, with an eccentricity of $0.4$.

In this run most of the mass in the disk actually accreted onto the
Jupiter mass planet. Part of the reason for this is that the smaller
planets were started at small radii, where the escape velocity from
the system exceeded the escape velocity from the Jupiter mass planet;
in that case we expect that most planetesimals will accrete onto
Jupiter rather than be ejected from the system. Only one body was
ejected. On the other hand, no bodies hit the star.

The latter result is representative of most of our runs; the fraction
of mass accreting onto the star is small, $\sim5\%$. The fraction
ejected varies with the initial semimajor axis and the mass of the
Jupiter mass body; both larger initial $\ap$ and larger $\mp$ produce
a larger fraction of ejected bodies (relative to bodies accreted onto
the massive planet). These rather low accretion fractions are in stark
contrast to those found by \cite{qh}. The difference appears to be
that we employ massive planetesimals, while their simulations employed
only test particles, which did not interact with each other. In our
simulations only the second most massive body (the first being
``Jupiter'') remains for long in a resonance; this second most massive
body lords it over his smaller brethren, kicking them out of nearby
resonances they might like to occupy. This tends to prevent the
smaller objects from reaching the extremely high eccentricities
($>0.9$) needed to strike the star.

In other runs the final state includes two planets in a mean motion
resonance. Since we start with such low planetesimal masses and
numbers, the mass ratio was always large. However, we expect that if
we allow larger terrestrial bodies to grow, that we may well find
final states with mass ratios nearer to unity. Finally, we note that
recent simulation of planetesimal migration show that two Jupiter-mass
object placed in a planetesimal disk will on some occasions migrate
toward each other \cite{hansen}. This could lead to resonance
capture followed by inward migration. Interactions between the massive
bodies and the planetesimal disk would likely tend to damp the
eccentricity of both bodies, but equation (\ref{both}) indicates that
as long as the outer body lost energy at a higher rate, the
eccentricities of both bodies would grow.

\section{DISCUSSION}

The mechanism we have proposed for the growth of eccentricity with
inward migration is essentially the same as that used to explain the
non-zero eccentricities of the inner Jovian satellites. It is well
understood and quite robust. It does not rely on the details of the
migration mechanism; in the case of the Jovian satellites the
migration is a result of the tidal bulge raised by Io on Jupiter; the
bulge exerts a torque on Io which transfers energy from Jupiter's spin
to the orbit of Io. Io in turn exerts, through a $2/1$ mean motion
resonance, a torque on Europa. In the satellite case the eccentricity
damping, produce by tidal flexing in both satellites as they oscillate
from peri- to apo-Jove, is very strong. This limits the eccentricity
to a value which, while small, is sufficient to dissipate enough
energy to power the volcanism on Io.

In previous sections we have examined resonant eccentricity growth in
the context of planetesimal migration, but it can work in the context
of migration due to tidal torques imposed by a gas disk as
well. Suppose that two Jupiter-mass bodies embedded in a gas disk are
locked in a mean motion resonance. Suppose that the gas between the
planets is removed, as numerical integrations indicate \cite{bryden}.
The gas inside the orbit of the inner planet will accrete onto the
star, possibly with some fraction being removed by a disk or stellar
wind. The planets are likely to follow the inner disk inward; if they
do not, the normal viscous spreading of the inner disk would move the
outer edge of the disk outward, until it experiences tidal torques
from the inner planet; this interaction would produce a back reaction
which would tend to damp the eccentricity of that planet; large
planetary eccentricities are unlikely to arise in that case.

However, it may be possible that the inner disk drains onto the star,
leaving the planets behind. This could occur, for example, if the
outer disk had a mass only slightly larger than that of the planets
\cite{npmk}. Then only the outer planet will experience significant
tidal torques, since the first order resonances of the inner planet
lie in the region between the planets that is depleted of gas.  Both
planets will then migrate inward, and eccentricity of the inner planet
will grow, since it does not experience much eccentricity
damping. This is exactly analogous to the planetesimal migration
described above, and the expressions we have given will describe the
growth and equilibrium values of the eccentricity once the appropriate
eccentricity damping rate for the outer planet is introduced (see,
e.g., Goldreich \& Tremaine 1980).

The tidal torque scenario also requires that the outer planet
have a mass sufficient to open a gap in the gas disk. If it does not,
then both bodies will experience tidal torques, which tend to damp
eccentricity rather strongly. An approximate criterion for gap
formation is \cite{lp86}
\be \label{simple_gap}
{m\over M_*}\gta 40\alpha (c_s/v_k)^2,
\ee 
where $\alpha$ is the the \cite{ss} viscosity parameter, $c_s$ is the
sound speed in the gas disk, and $v_k$ is the Keplerian rotation
velocity. If the disk is ionized, then the Balbus-Hawley instability
\cite{bh91,hb91} is likely to produce a rather larger (effective)
$\alpha$, of order $0.5$. The planet must then have a mass of order
$200$ Jupiter masses ($m_J$) in order to open a gap.  At small orbital
radii $a\lta 0.1$ AU the disk will be ionized \cite{gammie}. This
suggests that if the capture into resonance occurs at very small
radii, or if the planets migrate to very small radii, the eccentricity
of both bodies will be damped.

However, at larger radii protoplanetary disks are believed to be
substantially neutral, so that they are not subject to the
Balbus-Hawley instability \cite{gammie}. If so, they will likely have
a small effective viscosity, and equation (\ref{simple_gap}) predicts
that gap opening will occur for small (subJovian) mass planets. In
terms of $\alpha$ currently favored values are in the range
$\alpha\approx10^{-4}$ to $10^{-2}$. The latter value yields a mass
for a gap clearing planet of about $2m_J$. Smaller values of $\alpha$
would yield smaller masses, but it is believed that in that case a
second criterion is relevant, namely $m/M_*>3(c_s/v_k)^2$
\cite{pl84}. At $5$ AU this yields $m\gta m_J$. Thus the eccentricity
of the inner planet will only be excited if both objects are of
roughly Jupiter mass, assuming the migration is driven by tidal
torques in a gas disk.

Another constraint is that the migration torque not exceed the
resonant torque. If it does, the resonance will be broken, and the
eccentricity of the outer body will drop. The outer body may then
migrate inward, perhaps to be caught into a stronger resonance. This
constraint is likely to be important in the capture phase,
particularly in a migration produced by tidal torques. In that
situation, both planets are likely to have very small
eccentricities. The tidal torque is given by  \cite{ward}
\be \label{disk}
T_{disk}\approx \left({GM_*\mp\over 2a}\right){\mp\over M_*}
{M_{disk}\over M_*}\left({v_k\over c_s}\right)^3,
\ee 
assuming that the outer planet does not open a gap. For a disk of mass
$M_{disk}=10^{-2}M_*$ and a planet at $5$ AU, this is about
$50(GM_*\mp/2a)$.  If the outer planet is massive enough to open a gap, the
torque is set by the viscosity in the gas disk,
\be 
T_{gap}\approx\left({GM_*\mp\over 2a}\right)\alpha \left({c_s\over
v_k}\right)^2.
\ee 
This is much smaller than the torque in the gapless case,
$T_{gap}\approx10^{-5}(\alpha/10^{-2})(GM_*\mp/2a)$.
The resonant torque is
\be 
T_{res}\approx \left({GM_*\mp\over 2a}\right){m_1\over \mp}\left(
{ea_1\over a_1-\ap}\right)^q,
\ee 
where $e$ is the larger of the eccentricities of the two planets. This
eccentricity is likely to be small; if we take $e\approx 0.01$ then
the resonant torque is $T_{res}\approx 10^{-2}(m_1/\mp)(GM_*\mp/ 2a)$
for the $2/1$ first order resonance; near the inner edge of the gap
(where $a/\Delta a\approx 10$) this will rise by about 10.

If both bodies are of roughly Jupiter mass, capture is possible into
first or second order resonances, that is, resonances with
$|j_1-\jp|=q$, where $q=1$ or $2$, since $T_{gap}$ is the
appropriate torque to use.  This case may arise in the scenario
mentioned above, where gas caught between two giant planets can leak
out over several hundred orbital periods. The tidal torques from the
gas inside the inner planet will then tend to push it outward, while
the gas outside the outer planet will tend to push it inward; capture
into the $2/1$ or possibly the $3/1$ mean motion resonance could then
occur.

If the outer body has a mass substantially smaller than $m_J$, it will
not open a gap in the gas disk. If it has a mass comparable to or
larger than $1M_\oplus$, the hydrodynamic drag it experiences will be
much smaller than the tidal torques. It will undergo rapid (Type I in
the notation of Ward) inward migration, easily passing through any
mean motion resonances (see equation \ref{disk}). According to Ward,
the time for this inward migration will be less than $10^5$ years. The
inward migration will not halt until the outer body enters the gap
produced by the inner, Jovian mass object. It will then experience a
torque similar to that felt by the inner, Jupiter mass body, and both
bodies will migrate inward without a substantial change in their
eccentricity.

Since the initial inward migration is so rapid, it seems unlikely that
an outer planet with initial $m_1\lta 10M_\oplus$ will be able to
accrete sufficient solid material to trigger the accretion of gas
before it enters the gap produced by the inner planet. Once it enters
the gap, the outer planet could grow by eating other, inward migrating
bodies, a la the scenario proposed by Ward (1997) for explaining the
very short period Jupiter mass objects. However, unlike Ward's case,
the outer planet cannot emerge far enough from the inner edge of the
outer disk that its $2/1$ resonance leaves the disk, slowing the
inward migration; the inner planet is in the way. Given the mismatch
between the tidal and resonant torques, it seems likely that the
smaller planet will be subsumed by the more massive body.

There may be ways to distinguish migration by tidal torques and
migration by ejection of planetesimals. Planetesimal migration is
likely to produce dynamically isolated (although possibly not
unaccompanied) Jupiter mass bodies in small, eccentric
orbits; as we have seen, the small mass inner body responsible for
driving the eccentricity up to large values is often ingested into the
star or the Jupiter mass object.

In those cases where the inner body survives the migration process, it
should be possible to detect it with high precision radial velocity
observations. In some cases the inner body may have a large mass,
since planetesimal migration involving more than one Jupiter mass
object sometimes produces a convergence of the semimajor axes of two
bodies \cite{hansen}. This might produce systems like those recently
discovered around GJ876 \cite{876}. Alternately, such a system could
be the outcome of the migration of two Jupiter-mass planets in a gas
disk. 

Resonant migration in a gas disk is  less
likely to produce a single body in a moderately eccentric orbit than
is migration in a planetesimal disk; in the former case
both bodies are likely to be deep in resonance, and hence protected
from close encounters and the subsequent carnage. If they do suffer
close encounters, merger rather than ejection or accretion onto the
central star is the likely result. In a planetesimal migration, the
low mass of the inner planet combined with the frequency of close
encounters with numerous smaller bodies tends to keep the amplitude of
libration large. We have seen several cases in our numerical
integrations where the inner planet collides with the Jupiter mass
body, or with the central star.

There are several systems which have low mass planets in highly
eccentric orbits, including HD108147 ($M\sin i=0.35M_J$, $a=0.098$ AU,
$e=0.56$, and $K=37$m/s), HD83443 ($M\sin i=0.17M_J$, $a=0.174$ AU,
$e=0.42$, and $K=14$m/s) (the system contains at least two planets,
currently not dynamically linked), and HD16141 ($M\sin i=0.22M_J$,
$a=0.351$ AU, $e=0.28$, and $K=10.8$m/s). The first system is
particularly interesting as a test of the type of migration involved,
assuming the eccentricity is produce by resonant migration. The
putative resonant planet must have a mass less than about $1/3$ the
mass of the detected planet, in order to escape detection (since the
survey is clearly capable of finding planets with $K\approx 10$ m/s);
this would give $M\sin i\approx0.1M_J$. For a typical inclination we
would expect a mass of about of about $60M_\oplus$. Whether such a low
mass object could open a gap in a gas disk is an interesting question.
We note that the planet in HD108147 is near the radius at which the
Balbus-Hawley instability is believed to operate. In a gas disk, a
planet in this region would be subject to rapid eccentricity
damping. As radial velocity surveys improve below the $10$m/s level,
the discovery of even lower mass objects in eccentric orbits would
indicate that some mechanism other than resonant migration in a gas
disk was operating to produce the high eccentricities.

The fact that our simulations show accretion of planetesimals onto the
star suggests another way to distinguish the two scenarios. The
planetesimal migration is inevitably accompanied by the accretion of
$\sim5\%$ of the planetesimal disk mass onto the star; we see this even
in simulations in which we halt the migration at large semimajor
axis. Since the mass of the disk is of order $300-600M_\oplus$, this
amounts to $\sim20M_\oplus$ of rocky material accreted onto the
star. This material will include $5-7M_\oplus$ of iron, altering the
apparent metallicity of the parent stars dramatically. Note that
Jupiter contains only about $2M_\oplus$ of iron.

Moderate period (longer than 40 day period) systems produced by gas
migration are unlikely to pollute their parent stars so dramatically;
there is no reason to expect that such systems have a few hundred
Earth masses of rocky material lying around. The parent stars could
accrete metal rich Jupiter mass bodies after reaching the main
sequence; but again there is no reason to expect that every moderate
period system did so, when it is known that most stars that lack such
companions did not \cite{mcahn}. We say this because planets with 40
day or longer periods and moderate eccentricities ($\sim0.5$ or less)
are dynamically uncoupled from very short period planets (four days or
so), and so are unlikely to cause them to fall onto the star.

\section{CONCLUSIONS}
We have described how two resonant planets undergoing inward migration
can reach eccentricities of order $0.7$. The eccentricity growth is
largest when the inner planet is not subject to eccentricity
damping. Such a situation may arise either in planetesimal migration,
or in migration driven by tidal torques in which the inner gas disk
has been removed by accretion or mass loss in a wind.  We have
presented expressions for the equilibrium eccentricity, when it
exists, and for the relation between $e$ and the initial and final
semimajor axis of the resonant planets. 

We have presented numerical integrations showing that in some cases a
planetesimal migration will produce a single Jupiter mass object with
a large eccentricity; in other cases the Jupiter-mass object may be
accompanied by a resonant object with a similar mass, or by a
Neptune-mass companion. In the case of a Neptune-mass body the inner
companion will have a very large eccentricity.

We have also described integrations involving $10$ or more roughly
Earth mass bodies, together with a Jupiter mass planet; the latter is
forced to migrate inward, with the migration process tending to damp
its eccentricity. It typically captures one or more of the less
massive bodies into mean motion resonance. Usually only the second
most massive planet (which may not be the body that was originally
second most massive behind ``Jupiter'') survives long in resonance. In
fact, this second most massive body tends to accrete its smaller
companions.

We have suggested two possible ways to tell the difference between
systems produced by planetesimal or gas migration. First, if there
exist a large number of systems having an eccentric planet with either
no resonant companion, or with only very low mass (no-gap opening)
companions, a finding requiring very high precision radial velocity
measurements, it would strongly suggest that resonant migration by
tidal torques was not responsible. Second, the planetesimal migration
picture predicts that several Earth masses of iron will be accreted in
the migration process, well after the central star has reached the
main sequence. The tidal torque scenario is mute regarding this point;
accretion of material after the gas disk vanishes, and closely
correlated with the presence of Jupiter-mass planets, is not a natural
feature.

\appendix

\section{Appendix}

To derive the expression for the variation of the eccentricities of
two planets caught in a mean motion resonance when one or both are
subject to dissipative forces, we examine the equations describing the
evolution of energy and angular momentum, subject to the constraint
that the planets are locked in a resonance \cite{lpc, gomes98}.

We start with the energy, which is given by
\be 
E=-{G\ms m_1\over 2a_1}-{G\ms \mp\over 2\ap}.
\ee 
We assume $a_1<\ap$. It will prove useful to employ the variables
$\L_i\equiv\sqrt{GM_*a_i}$; then the resonance condition (\ref{lock})
implies
\be \label{action}
{d\ln \L_1\over dt}={d\ln {\cal L}_p\over dt}.
\ee 
This equation is accurate only in an average sense; on times shorter
than the libration time of the resonance it is violated.

We assume that some dissipative force removes both energy and angular
momentum from the orbits;
\be 
{dE\over dt}=\left({dE_1\over dt}\right)_T+
\left({d\Ep\over dt}\right)_T,
\ee 
where the subscript $T$ (for ``tides'') represents the effect of the
non-conservative force. After some algebra, the energy evolution
equation yields
\be 
\mp{d{\cal L}_p\over dt}+{\jp\over j_1}m_1{d\L_1\over dt}=
\mp\left({d{\cal L}_p\over dt}\right)_T + 
{\jp\over j_1}m_1\left({d\L_1\over dt}\right)_T .
\ee 
Combining this with equation (\ref{action}) we find
\be \label{appendix_energy}
{d\ln \L_1\over dt}={d\ln {\cal L}_p\over dt}={1\over
1+\left(m_1/\mp\right)\left(\jp/j_1\right)^{2/3}}
\left[\left({d\ln {\cal L}_p\over dt}\right)_T+
{m_1\over \mp}\left({\jp\over j_1}\right)^{2/3}
\left({d\ln \L_1\over dt}\right)_T\right] 
\ee 

Next we examine the angular momentum
\be 
L=m_1\sqrt{G\ms a_1(1-e_1)^2}+\mp\sqrt{G\ms \ap(1-\ep)^2},
\ee 
which evolves according to
\be 
{dL\over dt}=\left({dL_1\over dt}\right)_T+\left({dLp\over
dt}\right)_T.  \ee 
Introducing the auxillary variables $Y_i=\sqrt{1-e_i^2}$ we find
\begin{eqnarray} \label{angular} 
\lefteqn{\left[{dY_1\over dt}-\left({dY_1\over dt}\right)_T\right]
+{\mp\over m_1}\left({\jp\over j_1}\right)^{1/3}
\left[{d\Yp\over dt}-\left({d\Yp\over dt}\right)_T\right]} \nonumber\\
&&=Y_1\left[\left({d\ln \L_1\over dt}\right)_T -{d\ln \L_1\over dt}\right]
+{\mp\over m_1}\left({\jp\over j_1}\right)^{1/3}
\Yp\left[\left({d\ln {\cal L}_p\over dt}\right)_T -{d\ln {\cal L}_p\over dt}\right].
\end{eqnarray} 
Combining equations (\ref{appendix_energy}) and (\ref{angular})
gives (Gomes 1998)
\begin{eqnarray} 
\lefteqn{\left[{dY_1\over dt}-\left({dY_1\over dt}\right)_T\right]
+{\mp\over m_1}\left({\jp\over j_1}\right)^{1/3}
\left[{d\Yp\over dt}-\left({d\Yp\over dt}\right)_T\right]} \nonumber \\
&&=
\left\{
\left({d\ln \L_1\over dt}\right)_T-\left({d\ln {\cal L}_p\over dt}\right)_T
\right\}
\left[Y_1-{\jp\over j_1}\Yp\right]
\Bigg/
\left[
1+{m_1\over \mp}\left({\jp\over j_1}\right)^{2/3}
\right].
\end{eqnarray} 

Writing this in terms of $e_i$ and $a_i$, we find
\begin{eqnarray} 
\lefteqn{
{e_1\over \sqrt{1-e_1^2}}\left[{de_1\over dt}
-\left({de_1\over dt}\right)_T\right] 
+{\mp\over m_1}\left({\jp\over j_1}\right)^{1/3}
{\ep\over \sqrt{1-\ep^2}}\left[{d\ep\over
dt}-\left({d\ep\over dt}\right)_T\right] } \nonumber \\
&&=\left\{
\left({d\ln a_1\over dt}\right)_T-\left({d\ln \ap\over dt}\right)_T
\right\}
\left[
{\jp\over j_1}\sqrt{1-\ep^2}-\sqrt{1-e_1^2}
\right]
\Bigg/2
\left[
1+{m_1\over \mp}\left({\jp\over j_1}\right)^{2/3}
\right].\label{both}
\end{eqnarray} 

\acknowledgements Support for this work was provided by NSERC of
Canada.

\clearpage

\clearpage

\begin{thebibliography}{}

\bibitem[Balbus \& Hawley 1991]{bh91}
Balbus, S.\ A.\ \& Hawley, J.\ F.\ 1991a, \apj, 376, 214 

\bibitem[Bryden et al. 2000]{bryden}
Bryden, G., R{\'o}{\.z}yczka, M., Lin, D.\ N.\ C., \& Bodenheimer, P.\
2000, \apj, 540, 1091  

\bibitem[Ford et al. 2001]{ford}
Ford, E. B., Havlickova, M. \& Rasio, F. A. 2001, to appear in Icarus (astro-ph/0010178)

\bibitem[Gammie 1996 ]{gammie}
Gammie, C.\ F.\ 1996, \apj, 457, 355 

\bibitem[Goldreich 1965]{goldreich}
Goldreich, P. 1965, \mnras, 130, 159

\bibitem[Goldreich \& Tremaine 1980]{GT}
Goldreich, P.\ \& Tremaine, S.\ 1980, \apj, 241, 425 

\bibitem[Gomes 1998]{gomes98}
Gomes, R. 1998, \aj, 116, 997

\bibitem[Gonzalez et al. 2001]{gonzalez} 
Gonzalez, G., Laws, C., Tyagi, S., \& Reddy, B.\ E.\ 2001, \aj, 121, 432 

\bibitem[Hansen et al. 2001]{hansen}
Hansen, B., Murray, N. \& Holman, M. 2001, \apj, submitted.

\bibitem[Hawley \& Balbus 1991]{hb91} 
Hawley, J.\ F.\ \& Balbus, S.\ A.\ 1991b, \apj, 376, 223 

\bibitem[Holman \& Murray 1996]{hm}
Holman, M. \& Murray, N. 1996, \aj, 112 1278 

\bibitem[Laughlin 2000 ]{laughlin}
Laughlin, G.\ 2000, \apj, 545, 1064 

\bibitem[Lin et al. 1996]{lbr}
Lin, D.\ N.\ C., Bodenheimer, P., \& Richardson, D.\ C.\ 1996, \nat,
380, 606  

\bibitem[Lin \& Papaloizou 1986]{lp86}
Lin, D. N. C. \& Papaloizou, J. C. B. 1986, \apj, 307, 395

\bibitem[Lissauer et al. 1984]{lpc}
Lissauer, J. J., Peale, S. J. \& Cuzzi, J. N. 1984, {\em Icarus}, 58, 159

\bibitem[Marcy et al. 2001]{876}
Marcy, G. W., Butler, R. P., Fischer, D., Vogt, S. S., Lissauer,
J. J. \& Rivera, E. J., 2001, \apj, submitted.

\bibitem[Murray et al. 2001]{murray}
Murray, N., Chaboyer, B. \& Noyes, R. W. 2001, \apj, submitted

\bibitem[Murray et al. 2001]{mcahn} Murray, N., Chaboyer, B., Arras,
P., Hansen, B. \& Noyes, R. W. 2001, \apj, to appear (preprint
astro-ph/0011530)

\bibitem[Murray et al. 1998]{mhht}
Murray, N., Hansen, B., Holman, M. \& Tremaine, S. 1998 {Science},
279, 69

\bibitem[Nelson et al. 2000]{npmk} 
Nelson, R.\ P., Papaloizou, J.\ C.\ B., Masset, F. \& Kley, W.\ 
2000, \mnras, 318, 18 

\bibitem[\"Opik 1976]{opik}
\"Opik, E.J. 1976 {\em Interplanetary Encounters} (Elsevier Scientific,
Amsterdam)

\bibitem[Papaloizou et al. 2001]{PNM}
Papaloizou, J.C.B., Nelson, R.P. \& Masset, F. 2001, \aap, 366, 263

\bibitem[Papaloizou \& Lin 1984]{pl84}
Papaloizou, J.\ \& Lin, D.\ N.\ C.\ 1984, \apj, 285, 818 

\bibitem[Quillen \& Holman 2000]{qh} 
Quillen, A.\ C.\ \&  Holman, M.\ 2000, \aj, 119, 397 

\bibitem[Santos et al. 2000]{santos}
Santos, N.\ C., Israelian, G., \& Mayor, M.\ 2000, \aap, 363, 228 

\bibitem[Shakura \& Sunyaev 1973]{ss}
Shakura, N. I. \& Sunyaev, R. A. 1973, \aap, 24, 373

\bibitem[Ward 1997]{ward}
Ward, W.\ R.\ 1997, \apjl, 482,  L211 

\bibitem[Wisdom 1980]{wisdom} Wisdom, J.\ 1980, \aj, 85, 1122 


\end{thebibliography}
\end{document}